\documentclass[manuscript]{acmart}

\usepackage{svg}
\usepackage{xspace}

\def\xit{XIT\xspace}

\setcopyright{acmcopyright}
\copyrightyear{2022}
\acmYear{2022}
\acmDOI{XXXXXXX.XXXXXXX}

\acmConference[TRAIT 2022]{Make sure to enter the correct
  conference title from your rights confirmation email}{July 2022}{Online}
\acmBooktitle{TRAIT 2022: Trust and Reliance in AI-Human Teams} 
\acmPrice{15.00}
\acmISBN{978-1-4503-XXXX-X/18/06}

\renewcommand{\citeA}[1]{\citeauthor{#1} \cite{#1}}

\begin{document}

\title[Are we measuring trust correctly in explainability research?]{Are we measuring trust correctly in explainability, interpretability, and transparency research?}

\author{Tim Miller}
\email{tmiller@unimelb.edu.au}
\affiliation{%
  \institution{School of Computing and Information Systems and the Centre for AI \& Digital Ethics, The University of Melbourne}
 \city{Melbourne}
  \state{Victoria}
  \country{Australia}
  \postcode{3010}
}

\begin{abstract}
This paper presents an argument for why we are not measuring trust sufficiently  in explainability, interpretability, and transparency research. Most studies ask participants to complete a trust scale to rate their trust of a model that has been explained/interpreted. If the trust is increased, we consider this a positive. However, there are two issues with this. First, we usually have no way of knowing whether participants should trust the model. Trust should surely decrease if a model is of poor quality. Second, these scales measure perceived trust rather than demonstrated trust. This paper showcases three methods that do a good job at measuring perceived and demonstrated trust. It is intended to be starting point for discussion on this topic, rather than to be the final say. The author invites critique and discussion.
\end{abstract}


\maketitle

\section{Introduction}
\label{sec:intro}

\begin{comment}
\begin{tabular}{lp{11cm}}
\textbf{Collaborator:} & ``Unfortunately, the results showed that our explainability methods did not increase the participants' trust ratings of the model."\\

\textbf{Me:} & ``Should it have? Is the model itself trustworthy? If it is not very good, our explainability method should \emph{decrease} the trust, right?"\\

\textbf{Collaborator:} & ``Yes, I guess so. But our model performs quite well."\\

\textbf{Me:} & ``But how does the participant know this? Is our idea of `quite well' the same as theirs? I argue that there is not much point in us dwelling on these trust results unless we can establish whether it is trustworthy for these participants."\\
\end{tabular}

The above is a rephrasing of a conversation I had with a student of mine; and I've had several others similar. The conversation tries to draw out a common problem in evaluation of explainability, interpretability, and transparency (\xit) methods. As researchers, we propose a new \xit method, run some good user studies, and then see if the `trust' increases due to our new method. However, a good \xit method should not increase trust for the sake of it. If the underlying model we are explaining is a terrible model, a good \xit method should \emph{decrease} trust, not increase it. The ways most of us  are currently measuring trust do not seem to account for this.

This paper outlines why I believe most of us are not measuring trust correctly when we are assessing \xit methods, and shows some examples from the literature that I believe do a better job.
\end{comment}

As \citeA{hoffman2017taxonomy} notes, the idea of asking whether someone trusted their computer was once considered a strange question not worthy of asking. Why would people trust computers? As we have delegated more tasks to machines, trust in these machines has become of interest in areas from computer science to human factors. \citeauthor{Parasuraman1997-th}'s \shortcite{Parasuraman1997-th} seminal article on the use, misuse, disuse, and abuse of automation foreshadowed the problems of over- and under-reliance on machines due to lack of trust, while \citeA{Lee2004-fz} gave perhaps the first conceptual model of trust in machines and its processes that helped to clarify trust, reliance, and machines for many researchers.

In most research on trust and machines, trust is considered a mental attitude, such as a belief. The general hypothesis is that in most contexts, if a person trusts a machine\footnote{Throughout this paper,  the term `machine' is used a general term to describe computers, their software, individual applications, or even individual functions within applications.}, they will be more likely to rely on it. If they do not trust a machine, they will be less likely to rely on it, perhaps rejecting it entirely. Problems result when the alignment between whether a person trusts (distrusts) a machine does not align with whether they \emph{should} trust (distrust) a machine. As \citeA{Parasuraman1997-th} show, both under-reliance and over-reliance on machines can be problematic, causing issues such as physical, mental, or economic harm. One aim when building products and services is to engender \textbf{properly calibrated trust}: people trusting the parts that are trustworthy and distrusting the parts that are not trustworthy.

Over this period, researchers have also struggled with how to \textbf{measure trust}. Given that it is a mental attitude, this must be measured in field studies, lab experiments, and via surveys/interviews, with human participants. Much of this research has resulted in outputs like scales and surveys that ask study participants to rate various attributes of trust. 

However, these approaches measure \textbf{perceived trust}. While the perception of trust is an important thing to measure (it affects appropriation, adoption, and reliance), having participants merely state their trust is not the same as \textbf{demonstrating trust}. As such, other research has looked at how to measure trust via demonstration, such as the \emph{trust fall game} \cite{Miller2016-zv}, in which participants need to decide whether to use a particular agent to act on their behalf, demonstrating trust if they do. Further, even those these methods measure perceived trust, they are not designed to measure the effect of explainability, interpretability, and transparency (\xit) methods on the trust of participants.

In this paper, we look at methods to measure the \textbf{effect} that a \xit method has on trust. The key aspect that makes measuring this different to existing methods such as the trust fall game, is that we are not concerned with the trust of the underlying machine --- we are instead concerned with the effect of using different \xit methods, which we call \emph{interventions}. We discuss how to measure the effect of \xit methods on both calibrated trust, both perceived and demonstrated.

The key insight is that to measure whether \xit methods have an effect on calibrated trust, we must know whether the participant should consider underlying model being explained/interpreted is trustworthy. As this is not possible to know \emph{a priori}, trust evaluation methods should \textbf{manipulate} the trustworthiness of models and then measure whether the \xit methods are able calibrate trust (distrust) for more (less) trustworthy systems. 

The trust evaluation methods in this paper are not new -- they are taken from existing literature. This paper aims to: (a) make the problems of measuring trust for \xit methods better understood by the research community; (b) share trust evaluation methods with this community; and (c) start a discussion on other ways to approach the problem.

\section{Related Work}
\label{sec:background}

In this section, we give a high-level overview of  existing research on trust measurement.
\subsection{Measuring Perceived Trust}

Measuring perceived trust has received more attention than measuring demonstrated trust. Much of the work has focused on defining what constitutes trust to enough fine grain that we can measure the components of trust.

\citeA{Jian2000-kk} are the first researchers to empirically derive a questionnaire for human-human trust and human-machine trust. The questionnaire is mostly targeted towards automation/autonomy, given the terms used.
%There are two results that are commonly assumed but empirically derived by \citeauthor{Jian2000-kk}:
%\begin{enumerate}
%    \item  Trust and distrust are opposite ends of a continuum, rather that separate concepts.
%\item Human-machine trust is  similar to interpersonal human-human trust -- the empirical models of both that \citeauthor{Jian2000-kk} derive are almost the same. The authors conclude: ``This implies that people do not perceive concepts of trust differently across the different types of relationship.'' 
%\end{enumerate}
%
\citeauthor{Jian2000-kk} propose a trust checklist consisting of 12 Likert-scale based items, ranging from deceit, to reliability, to integrity.

Since then, other scales of trust have been proposed, often with specific use cases in mind; e.g.  \citeA{cahour2009does}, \citeA{wang2009trust}, \citeA{Wang2021-cv}. These scales are typically based on \citeauthor{Jian2000-kk}'s \cite{Jian2000-kk} original scale, and there is high overlap in the scale components, although \citeA{Wang2021-cv} propose a scale specific to data-driven AI methods.
\citeA{Dizaji2021-qr} conduct a comprehensive review of trust measurement scales, viewing them through the lens of the  IMPACTS model of trust \cite{Hou2021-fw}. Their main finding is that the component of \emph{adaptivity} is not well considered in trust and automation research.

Most relevant to this paper is the trust scale developed by \citeA{hoffman2018metrics}, which is a trust scale aimed at explainable AI. Like other scales, it is based on \citeA{Jian2000-kk}, with some items adapted from \citeA{cahour2009does} and \citeA{wang2009trust}.
However, while for use in explainable AI, the scale proposed by \citeA{hoffman2017taxonomy}, does not explicitly measure the effect of an \xit intervention, as it is a scale, not a process. This scale and others outlined in this section can be used in a larger evaluation process to measure the effect of \xit methods, as we outline in Section~\ref{sec:method}.

\subsection{Measuring Demonstrated Trust}

Measuring demonstrated trust has also received some attention in the literature. Later in Section~\ref{sec:method} we discuss work from \citeA{hussein2020trust}, \citeA{huber2021local} and \citeA{Schmidt2019-jy}, who propose and/or use measures of demonstrated trust.

The idea of demonstrated trust between people is seen in areas such as supply chains \cite{Laeequddin2010-mm, Tejpal2013-bl}, organisations \cite{McEvily2011-jb}, journalism \cite{Engelke2019-qp}, and online interactions \cite{Nurse2014-go}. But perhaps the most explicit research that looks at demonstrated trust is in economic game theory. The \emph{investment game} \cite{camerer1988experimental, berg1995trust} is a game in which participants are given real amounts of money and must `invest' it. The payoffs of the players are dependent on each other, not just their own actions. By varying the rules and amounts, researchers can determine the amount of trust between players. 
However, this work measures interpersonal trust between two or more people, rather than the uni-directional trust between a human and a  machine.

The \emph{trust fall game}, proposed by \citeA{Miller2016-zv}, aims to measure demonstrated trust in machines using reliance. That is, trust is demonstrated by relying on an agent. The game is based on the well-known `trust fall' activity, in which a person, the \emph{trustor} closes their eyes, falls backwards, and one or more people, the \emph{trustees}, catch them. Anyone who does not trust the people catching them will not submit themselves to this test. \citeA{Miller2016-zv} use this analogy for a trust fall game for machines. Participants in an evaluation first establish a mental model by observing or interacting with a machine. Then, participants' trust is then tested by giving them to choice to rely on the machine's answers to problem, or to solve it themselves. The hypothesis is that if the participant has deemed the machine to be trustworthy, they are more likely to rely on the machine than if they deem the machine to not be trustworthy.

The trust fall game is a nice model for evaluating \textbf{demonstrated trust}, however, on its own, it is not sufficient for measuring the effect of an intervention such as a \xit method. As there is no way to assert what level of trust a participant \emph{should} trust (or distrust) the agent, it is not possible to say if the \xit intervention is working as intended.
\section{Trust in Human-Machine Scenarios}
\label{sec:trust}

In this section, we present an existing definition of trust in human-machine scenarios, which we use as a basic for defining the parameters of good evaluation methods.

\begin{definition}[Interpersonal trust \citeA{Mayer1995-xt}]
\label{defn:interpersonal_trust}
\citeA{Mayer1995-xt} define \textbf{trust} as: ``the willingness of a party to be vulnerable to the actions of another party based on the expectation that the other will perform a
particular action important to the trustor, irrespective of the ability to monitor or control that other party.''
\end{definition}

\citeA{jacovi2021formalizing} build on this model by first acknowledging that human-machine trust is similar to directional, interpersonal trust, except where the trustee is a machine doing some action, such as providing advice, finding information, executing a command, etc.
\citeauthor{jacovi2021formalizing}'s definition goes further to consider two additional key elements: (1) the \emph{context} in which the action is to be performed; and (2) \emph{what} exactly is the machine being trust with.

The \emph{what} part in \citeauthor{jacovi2021formalizing}'s model insists that trust be considered specifically against a set of \emph{contracts}, which can be a legal or social agreements that document the expected behaviour of a machine for given tasks. For example, there may be a social expectation that a machine learning model can  accurately and fairly predict the right products to promote to people online, but not necessarily in an explainable manner. As another example, in an air control system, there may be a legal contract spelling out that calculations of collisions between aircraft must be correct, but also completed within a specific time limit. 

Bringing Definition~\ref{defn:interpersonal_trust} and the notions of context and contract, \citeauthor{jacovi2021formalizing} define human-machine trust as follows.

\begin{definition}[Human-machine trust and distrust \cite{jacovi2021formalizing}]
If and only if a person believes that a machine $M$ will uphold a contract $C$ in context $n$, and the person accepts vulnerability to $M$'s actions, then that person \textbf{trusts} $M$ contractually to $C$ in $n$. 

If and only if a person  does not accept vulnerability to machine $M$'s actions \emph{because} they believe that $M$ will fail to uphold a contract $C$ in context $n$, then that person \textbf{distrusts} $M$ to uphold contract $C$.
\end{definition}

\begin{definition}[Machine trustworthiness \cite{jacovi2021formalizing}]
A machine is \textbf{trustworthy} to a contract $C$ if and only if it is capable of maintaining this contract.
\end{definition}

Often, we see the conflation of trustworthiness with trust, or definitions of trust that are defined in terms of trustworthiness. \citeauthor{jacovi2021formalizing}'s definition above cleanly separates trust and trustworthiness. This allows us to then define the model in Table~\ref{fig:warranted-unwarranted-trust}. 
Inappropriately trusting when a something is not trustworthy is \emph{unwarranted} trust; and a similar concept exists for \emph{unwarranted distrust}. However, they mean that trust (and distrust) are \textbf{caused by} the ability to uphold a contract; or more succinctly, caused by the trustworthiness of the model, rather than being coincidental.

\renewcommand{\arraystretch}{2.0}
\begin{table}[!th]
\centering
\begin{tabular}{lll}
    & \textbf{Trustworthy} & \textbf{Not trustworthy}\\
\hline
\textbf{Trusted}     & Warranted trust & Unwarranted trust\\[1mm]
\textbf{Distrusted} & Unwarranted distrust & Warranted distrust\\[1mm]
\hline
\end{tabular}
\caption{Warranted and unwarranted trust}
\label{fig:warranted-unwarranted-trust}
\end{table}
\renewcommand{\arraystretch}{1.0}

From this model of warranted/unwarranted trust/distrust, we can clearly state that the goal with respect to trust should be \textbf{appropriately calibrated trust}. That is, we should aim to avoid unwarranted trust and unwarranted distrust, rather than just aim for trust irrespective of whether a machine is trustworthy for particular contracts/contexts.

%However, the \textbf{goal of trust} is not merely to have appropriately calibrated trust itself. Trust is important because it supports predictability in social interaction, which means that people can rely on others without having to take up cognitive load worrying about outcomes, etc. \cite{misztal2013trust}. This predictability is key to trust in both human-human relationships as well as to how people trust machines. For example, if we know under what circumstances we can trust a machine to find some information, we can rely on the resulting information without having to check it ourselves (or with minimal checking).

The \textbf{four important lessons} from this section as related to evaluating trust are:
\begin{enumerate}
    \item the \textbf{vulnerability} of the trustee to the trustor, which implies that the trustee is taking a risk by willingly allowing themselves to be vulnerable;
    \item the \textbf{expectation} of the trustor that the trustee will perform the action in the best interests of the trustee;
    \item the outcome of that action is important to the trustor -- they have a \textbf{stake} in the outcome; and
    \item warranted trust and distrust are caused by the trustworthiness of the trustee, and we should aim to \textbf{avoid unwarranted trust} as well as unwarranted distrust.
\end{enumerate}

One may argue that this is perhaps an over-simplified model of trust, and they would be right that it fails to capture much of the realities of trust in human societies and human-machine trust.  In reality, any type of trust is tentative and trust in machines is a dynamic process that changes over time as people interact with the machine. Further, it is not a binary concept, but a sliding scale, perhaps better represented by a probability or confidence. Nonetheless, we believe that this model is useful. We argue in the rest of this paper that in XIT research, we are failing to measure even such basic notions of trust.

\section{Requirements for Evaluating Warranted/Unwarranted Trust/Distrust}
\label{sec:requirements}

If we want to evaluate trust in field studies or lab studies that asses \xit methods, we argue that any evaluation must have the following requirements:

\begin{description}

\item [Task performance] The tasks done by participants in the evaluation, whether part of an experiment or prior to any discussion, must have some measurable performance. That is, participants must be completing some task that has an outcome. \textbf{Rationale}: The goal of trust is not simply to have trust, but to support predictability in social interaction.

\item [Risk] Subjects must be vulnerable to the risk of the tasks being evaluated. That is, there must be some downside to having unwarranted trust or unwarranted distrust in the evaluation. Further, the participants must be aware of the risk. \textbf{Rationale}: Trust cannot exist without vulnerability. \citeA{herbert1970conceptual} similarly argue that to study trust, any interaction must have risk, and the the participants in that interaction must be aware of them.

\end{description}

Further, to measure demonstrated trust, as opposed to (or in addition to) perceived trust, we have another requirement:

\begin{description}
\item [Reliance] Participants must be given an opportunity to choose whether to rely on the machine. \textbf{Rationale}: For trust to exist, people must \emph{accept} vulnerability to risk. Without having to rely on the machine in some way, there is no acceptance of the risk involved, and so no trust. 

\end{description}

Methods for explainability, interpretability, and transparency are interventions that aim to manipulate (usually positively) people's understanding of models. To evaluate trust of \textbf{interventions}, we have one further requirement:

\begin{description}

\item [Manipulated trustworthiness] There must be some known or estimated `level' of trustworthiness that is manipulated as part of the evaluation. Trust measures must be taken for these different levels. \textbf{Rationale}: If we want to measure \emph{warranted} trust (that is, caused by the trustworthiness), we cannot establish whether the intervention has correctly calibrated trust without manipulating the trustworthiness of the machine. For example, if we try technique A against baseline B, and technique A is shown to engender higher trust, this is meaningless if we do not know the trustworthiness. A good intervention should \emph{decrease} trust if the model is less trustworthy than the trustee initially believes. Given that participants in studies would have little-to-no evidence to establish trust initially, manipulating trustworthiness to different, known/estimated levels allows a before/after comparison. 
\end{description}

\section{Measuring Trust}
\label{sec:method}

In this section, we showcase three methods for measuring the effect of trust on \xit interventions. All of these methods are extension to the idea of the \emph{trust fall game} \cite{Miller2016-zv}, discussed in Section~\ref{sec:background}.

\subsection{Approaches to Measuring Trust}
\label{sec:method:approaches}

There are three broad categories of evaluation for measuring trust:

\begin{description}
    \item [Questionnaires or surveys] Participants in a study have access to some evidence of trustworthiness, and are asked to rate their opinion of trustworthiness. \citeA{hoffman2018metrics} present a survey instrument specifically for measuring trust within explainability (which in our view extends to any intervention intended to calibrate trust). It is important to emphasise: these measure participants' \emph{perception} of their trust. It is well known that self reporting in experiments can be inaccurate, and even in the are of trust, experiments show that people's reported ratings of trust do not correspond with their actions \cite{Miller2016-zv}. 
    
    \item [Interviews, focus groups, and similar reporting tools] Researchers interview people to ask about their experiences and how this influences trust. These offer much richer insights into the mental model of participants, but are more labour intensive, and like questionnaires and surveys, they give insight into perceived trust, rather than actual trust. 
    
    \item [Reliance] Participants are given the option to use or rely on something. The more often they choose not to rely on this, the less they are considered to trust it compared to the alternative. This demonstrates trust, rather than rating the perception of trust.
\end{description}

None of these methods is `the correct' or best way to measure trust. While reliance is a better measure of calibrated trust, the perception of trust is also important. One would not want to design an explainability technique the increases reliance, but the perception of users is that they do not trust it. Ideally, we want perceived trust to align with reliance; and ideally, we would take measures using all three. Whichever category a particular measurement tool falls into, the requirements outlined in Section~\ref{sec:requirements} must be considered if we want to gain proper insight into trust.

\subsection{Measuring the Effect of \xit Interventions on Trust}

In this section, we outline three protocols that can be used to evaluate trust in research where we are testing the effect of a \xit intervention on trust. These three protocols have been used by other researchers, and we identify papers where we first read about them.

\subsubsection{Within-subject design}
\label{sec:method:within}

\begin{figure}[!ht]
\centering
\includegraphics[scale=0.5]{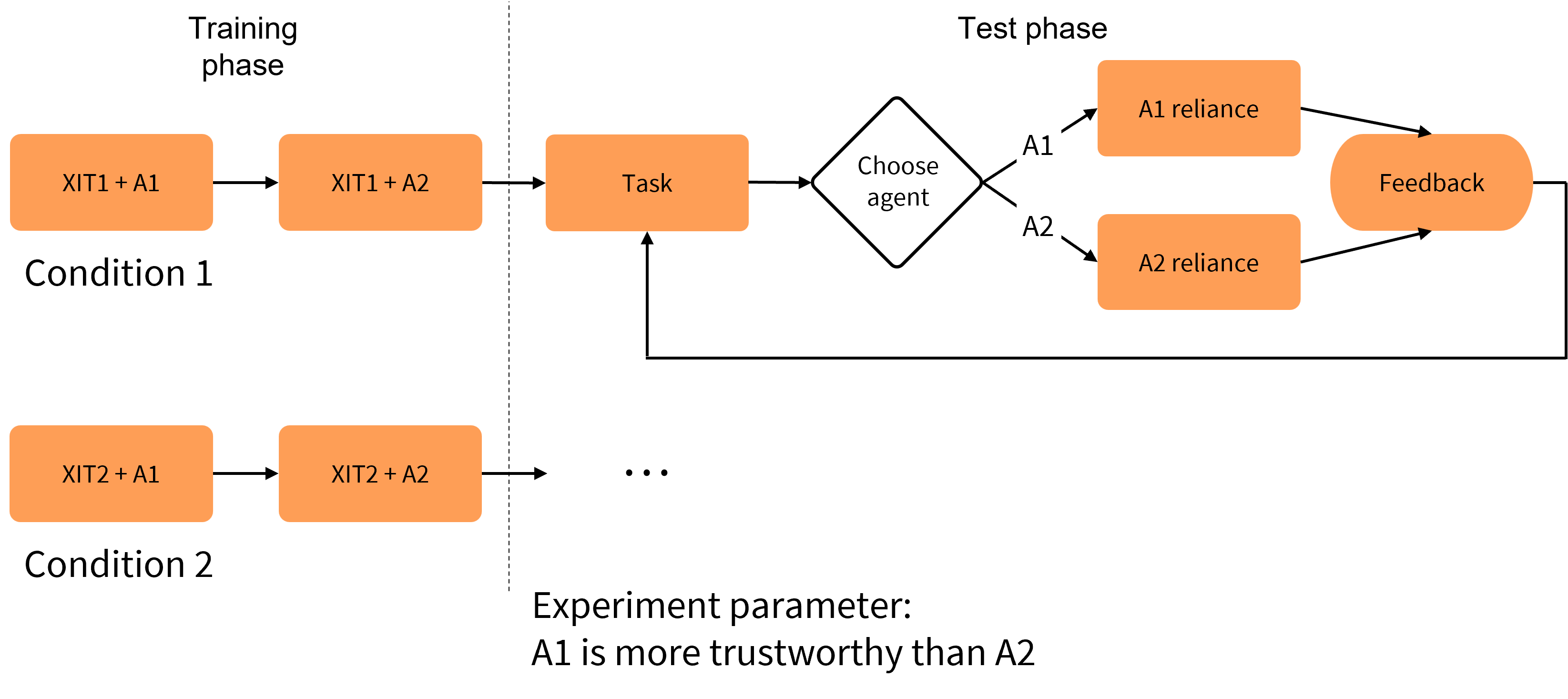}
\caption{A within-subject measure of demonstrated trust for \xit methods, used by \citeA{huber2021local}. Participants interact with two or more agents with differing levels of trustworthiness, and then must choose one to rely on to complete each task in a sequence of tasks. One condition for each method.}
\label{fig:within-subject-trust-measurement}
\end{figure}

Figure~\ref{fig:within-subject-trust-measurement} shows a within-subject design  used by \citeA{huber2021local} in their experiments measuring the effects of using saliency maps and summaries for explaining agent policies. 

In this design, each participant acquires some experience of two or more agents of differing levels of trustworthiness. This could be a training phase of an experiment or practical experience in the field. The differing levels of trustworthiness are known by the researchers, but not the participants. This can be achieved by e.g.\ inserting random noise to reduce the performance of a machine learning classifier. Participants also interact with an \xit technique (or baseline) to help determine the trustworthiness of the agents. To avoid learning biases, the order in which the participant interacts with the agents must be counter-balanced. 

In the test phase, participants must then choose between the two agents to do a task or series of tasks.  Importantly, the output of the agent and the \xit technique is not provided during this test phase. Participants must make a judgement whether to rely on the agent without seeing its answer. This reliance demonstrates trust.

If there are multiple tasks in the test phase (e.g. multiple scenarios in a focus group), averaging the number of times each agent was used allows us to then give an overall `score' for each agent. Given that agent A1 is more trustworthy than agent A2, the assumption of this design is that if an \xit technique is better for engendering warranted trust and distrust, then participants should be able to better determine the trustworthiness of agent A1. They would therefore rely on agent A1 more often than agent A2. Finally, a trust scale or survey can be administered to measure perceived trust.

One can see that this design adheres to our requirements from Section~\ref{sec:requirements}. First, there is a task that is completed throughout the study. Second, the participants are vulnerable to the risk of choosing the less trustworthy agent, provided this is part of the study (see Section~\ref{sec:method:vulnerability} for more details on this). Third, the participants are given the choice to rely on one of the agents. Finally, the trustworthiness is manipulated by the researchers.

\subsubsection{Between-subject design}
\label{sec:method:between-subject}

\begin{figure}[!ht]
\centering
\includegraphics[scale=0.5]{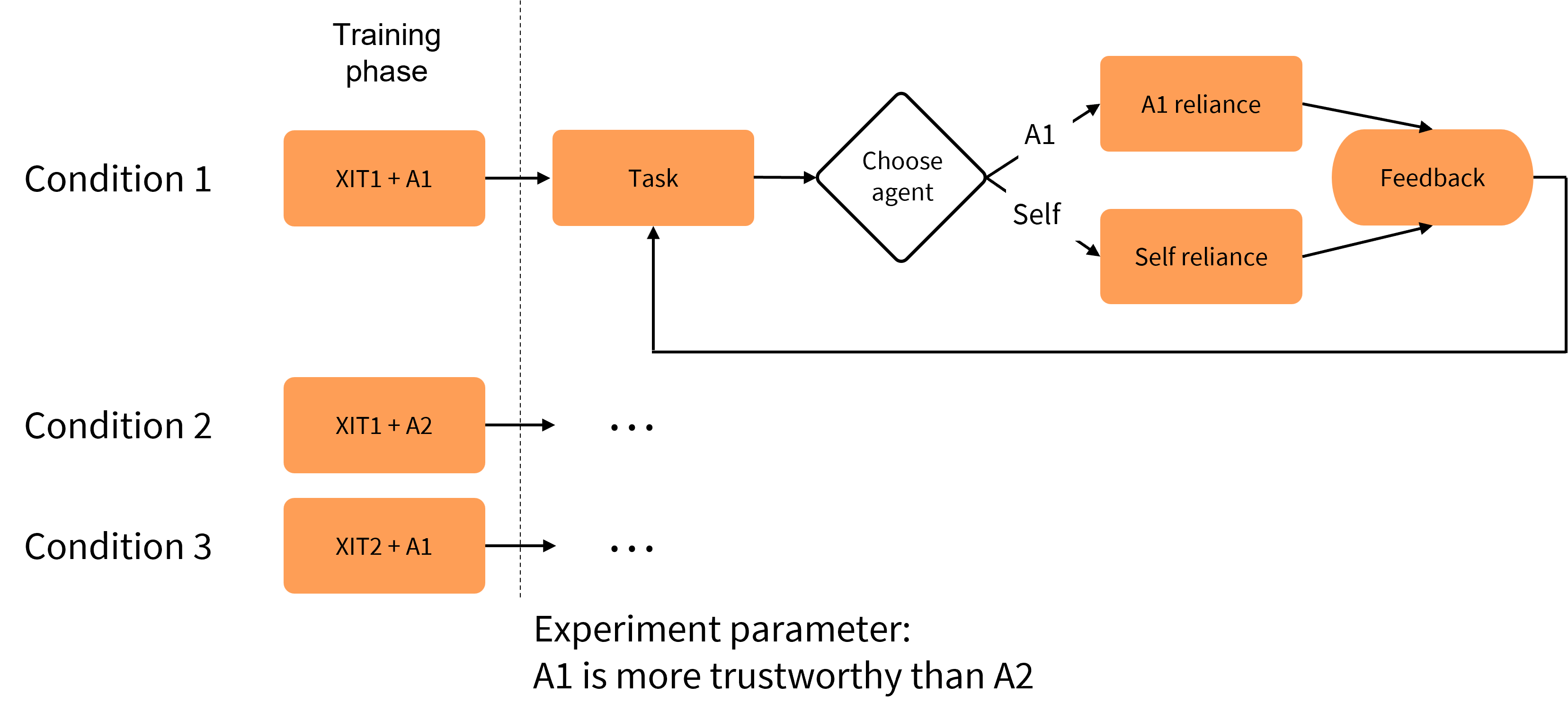}
\caption{A between-subject measure of demonstrated trust for XIT methods. Participants interact with an agent, and then must choose whether to rely on themselves or the agent to complete each task in a sequence of tasks.  One condition for each pair in method $\times$ agent.}
\label{fig:between-subject-trust-measurement}
\end{figure}

Figure~\ref{fig:between-subject-trust-measurement} shows a between-subject design for measuring trust. The rationale for this design is the same as the the within-subject design.
The basic design is similar to the within-subject design from Section~\ref{sec:method:within}, with `known' trustworthiness of two agents A1 and A2, and a choice. The difference is in two places:
\begin{enumerate}
    \item Each participant sees just one agent instead of two.
    \item The choice is between the agent and the participant performing the task themselves (self reliance).
\end{enumerate}

\citeA{hussein2020trust} use a cross between these two designs in their experiments measuring the effects of transparency on human-agent decision making in swarm systems. Each participant is tasked with two of the conditions from Figure~\ref{fig:between-subject-trust-measurement}: one for each agent to account for trustworthiness, with the order counterbalanced to avoid ordering effects. The choice is between the agent and the participant, rather than the two agents. The researchers also task participants to fill out a (perceived) trust questionnaire at intervals. 

The advantage of \citeauthor{hussein2020trust}'s design is that the researcher can perform mediation analysis on the relationship \emph{Trustworthiness} $\rightarrow$ \emph{Perceived Trust} $\rightarrow$ \emph{Reliance} to determine how much the trustworthiness affects perceived trust, which in turn affects reliance, versus how much the trustworthiness affects reliance (or actual trust) on its own.

\paragraph{Within- vs.\ between-subject design}

The main advantage of a between-subject design is any ordering effects are eliminated, rather than controlled for. However, a small advantage of a within-subject design, such as those used by \citeA{huber2021local} and \citeA{hussein2020trust} is that we can run post-experiment surveys or interviews where the participants can give their impressions on the \textbf{difference} between the trustworthiness of the agents. This direct comparison gives participants reference points against which to compare agents.

\subsubsection{Manipulating trust at the instance level}

\begin{figure}[!ht]
\includegraphics[scale=0.5]{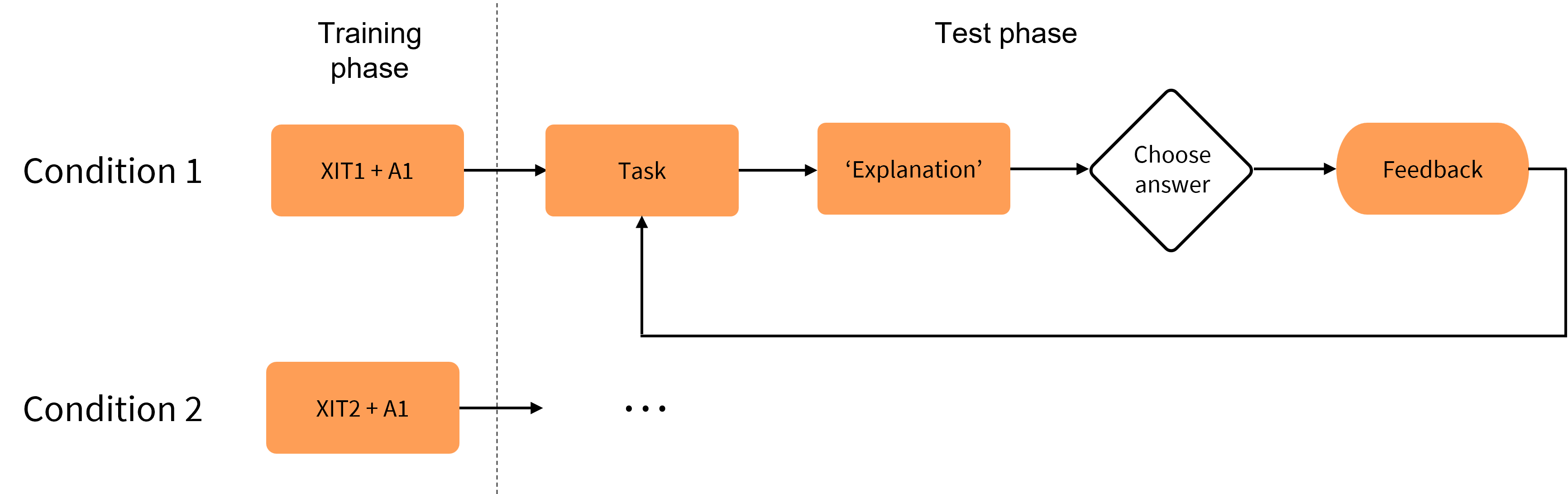}
\caption{A measure of trust for XIT methods based on information gain. Participants interact with an agent, and then must choose whether to rely on themselves or the agent to complete each task in a sequence of tasks.  One condition for each pair in method $\times$ agent.}
\label{fig:instance-level-manipulation}
\end{figure}

\citeA{Schmidt2019-jy} propose a different design in which the manipulation of trust is at the task or instance level. Rather than having different agents with different levels of trustworthiness, individual tasks or instances are chosen to be `high` or `low' quality. The overall framework is outlined in Figure~\ref{fig:instance-level-manipulation}. Instead of two or more agents, a single agent is used. The accuracy/performance of agent decisions in individual tasks is known; that is, there is a series of tasks/scenarios, and the ground truth of these is known. A trust score is calculated based on agreement between participant and agent. An important distinction between this design and the earlier two designs is that the participants see the `explanation' (or \xit method) instead. This hypothesis is that a better \xit method will give a better understanding of the domain.

\citeA{Schmidt2019-jy} first define the \emph{information transfer rate} (ITR) from an \xit method as:

\vspace{2mm}

\begin{tabular}{lll}
$ITR =  \frac{I(Y_h, Y_a)}{t}$ & st. &
$I(Y_h, Y_h) = \Sigma_{y_a, y_h} p(y_a, y_h) \log \frac{p(y_a, y_h)}{p(y_a) p(y_h)}$\\
\end{tabular}

\vspace{2mm}
\noindent
where $Y_h$ and $Y_a$ are the set of answer given by the human ($h$) and agent ($a$) respectively, $y_h \in Y_h$ and $y_a \in Y_a$ are individual answers respectively, and $t$ is the amount of time spent by the human participant answering the set $Y_h$.
So, \citeauthor{Schmidt2019-jy} define ITR using \emph{information gain}.  The more the human participant and the agent agree on answers, the higher the ITR. Time is used with the assumption that a better understanding provided by a \xit technique will enable faster answers. For the rest of this paper, we will not discuss $t$.

Trust is then defined as $T = \frac{ITR_{Y_a}}{ITR_Y}$, where $ITS_{Y_a}$ is the ITR between the agent and the human participant, and $ITR_Y$ is the ITR  between the participant and the ground truth.
So, trust is a fraction of the ITR of the agent+human over the ITR of the human+ground truth. A score greater than 1 indicates a participant who has `too much' unwarranted trust in the agent, while a score less then 1 indicates a participant who has `too much' unwarranted distrust. A score of 1 is `perfectly calibrated trust'.

However, this measures agreement rather than reliance, so is not truly measuring \emph{demonstrated} trust. Nonetheless, this approach can be modified to measure demonstrated trust: showing the task, withholding the explanation (and agent decision), and then asking the participant to choose between the agent and self reliance, as is done in Section~\ref{sec:method:between-subject}.

An advantage of manipulating trust at the instance level is that it gives us a finer-grained control to measure trust calibration. By choosing instances based on their alignment with the ground truth, we can measure the calibration by measuring the number of false positives and negatives.

A disadvantage of this approach is that it requires ground truth, which is often not available or not really `ground' at all. The previous approaches that manipulate trustworthiness by using different agents is easy to achieve by simply taking one agent and adding noise to their answers, as is done by \citeA{hussein2020trust}. 

\subsection{Vulnerability in laboratory experiments}
\label{sec:method:vulnerability}

`Vulnerability to risk' is a key component of trust. The designs above did not consider risk explicitly. 
In a questionnaire/survey or field study of trust, participants must be (or have been) vulnerable to the risk of the agent failing. Reliance without risk does not require trust.
In experimental situations, participants must also be have stakes in the situation. This is difficult because in many cases, participants know that after any experiment, they can return to  their normal lives with little impact. In such cases, we need to provide incentives that simulate risk.

There are two ways we have seen that can introduce risk:

\begin{description}
    \item[Payment bonuses] As noted earlier, because a good design requires that participants undertake some measurable tasks, we can provided bonuses on top of the base payment to participants who excel. For example, a small bonus per task that is completed well, or a single bonus for reaching a particular milestone.
    In some cases, we may want to measure the trust impact on participants playing the role of \emph{decision subjects} --- that is, simulating situations in which a \emph{decision maker} uses an algorithm to make a decision about a decision subject, such a loan decision being made about a borrower. In this case, decision subject has no control over the outcome. In these cases, decision subjects receive bonus payments depending on the behaviour of other participants. This is less straightforward from an ethical perspective, but is not uncommon in experiments of bi-directional trust (see Section~\ref{sec:background}; and is a further reason why risk bonuses must be on top of the base payment for participants.
    
    The reader may question whether missing out on bonus payments can be considered risks; and it is true that we cannot compare the risk losing of a small financial bonus with the risk of e.g.\ undergoing an invasive medical treatment. However, \citeA{Bradler2019-xy} show that financial bonuses for performance increased output in both a creative task and a simple task; and studies such as those by \citeA{Amir2012-bx} show that small bonuses affects people's decisions in economic games, even when played in online platforms like Amazon Mechanical Turk, where participants are quite anonymous to each other. So, we argue that participants are presented with a risk and they are vulnerable to this risk. As a result, we are  \textbf{measuring trust} when we use financial bonuses for performance.  However, we should be careful to not over-interpret the results as saying that our models are trustworthy for high stakes decisions when the laboratory stakes are low.
    
    \item[Gamification] Tasking participants to play games with measurable performance can increase people's engagement and avoid player fatigue. \citeA{Guttman2021-fy} argue that this makes them an good proxy for measuring human-AI interactions, compared to other synthetic tasks that are not gamified, because they ``allow us to study how framing affects human-AI interfaces in more realistic ways than laboratory experiments alone".
    
    We argue that the risk of losing points in a gamification environment is a reasonable proxy for low stakes risk; enough to measure trust as reliance. 
    For example, in a laboratory experiment, we used gamification to increase participant engagement. Outside of the laboratory, some participants created a leaderboard. Once we have a handful of names on the leaderboard, people started asking if they could do the experiment again for no payment, so they could work their way up the leaderboard.
    %
    %As another example, we are surely all aware of the game Wordle that exploded onto our social media feeds in late 2021. Not only does the gamification aspect of Wordle mean that we want to perform well for ourselves (and don't you hate it when you make a mistake!), the trend of sharing our results on social media means that we have some reputational risk of doing poorly. This kind of reputational risk is likely acceptable in experimental conditions because the stakes are fun rather than harmful.
    
    As with financial bonuses, the risk is low, so we should be careful how we interpret results from games when the stakes are low. 

    \item[Deception] A third approach is to deceive participants into thinking that they are at risk. There are clear ethical concerns with such an approach, but if handles ethically, can be a powerful method.
    
\end{description}
\section{Discussion}
\label{sec:discussion}

It is important to note that the evaluation methods in this paper are not restricted to laboratory experiments: they generalise  to field studies. In addition, the general design of intervening on trustworthiness is required by all three types of measuring trust outlined in Section~\ref{sec:method:approaches}: questionnaires/surveys, interviews/focus groups, and reliance. 

All three evaluation methods have some intervention on trustworthiness --- either on the agent itself or on individual decisions. Can these approaches  be used in field studies where we cannot manipulate trustworthiness? We believe they cannot; however, this is not a weakness of the approaches -- it is the reality of measuring the effect of \xit methods on trust. Without manipulating trustworthiness, we are not truly measuring trust.

Note that the idea of manipulating trust is not only important for demonstrated trust, but also for perceived trust. As with the approached outlined in this paper, if we want to measure whether our \xit method \emph{correctly} impacts the perceived trust of participants, we need to manipulate the trustworthiness of the underlying agent.

One final point to make is that trust is a dynamic concept, and people's trust in machines varies over time \cite{hoffman2017taxonomy}, even in the presence of \xit methods. As \xit researchers, we need to start evaluating the impact of \xit methods in more longitudinal studies, measuring trust as well as other crucial factors. That is not to say that such longitudinal studies are easy -- they are difficult. However, any such studies will be highly valuable to the community.

This position paper is not intended to be the end of the conversation. We encourage people to critique these methods, propose changes, and propose new methods. The primary reason for this paper is to promote discussion and make the research community aware of existing work on measuring the effect of \xit interventions on trust.

\begin{acks}
Thanks to Alon Jacovi, Ana Marasovic, and Yoav Goldberg for their discussions about trust and trustworthiness.  

This research is supported by Australian Research Council
(ARC) Discovery Grant DP190103414: \emph{Explanation in Artificial Intelligence: A Human-Centred Approach}.
\end{acks}

\bibliographystyle{ACM-Reference-Format}
\bibliography{references}

\end{document}